\documentclass[11pt]{article}%
\usepackage{amsmath}
\usepackage{graphics}
\usepackage{color}
\usepackage{epsfig}
\usepackage{graphicx}%
\usepackage{amsfonts}%
\usepackage{amssymb}
\usepackage{setspace}

\newtheorem{theorem}{Theorem}[section]
\newtheorem{definition}{Definition}[section]

\newtheorem{lemma}{Lemma}[section]

\newtheorem{proposition}{Proposition}[section]
\newenvironment{proof}{\noindent{\bf Proof.}}{  \rule{2mm}{3mm}}

\date{}

\allowdisplaybreaks
\begin{document}

\title{The 2-valued case of makespan minimization with assignment constraints\thanks{This research has been co-financed by the European Union (European Social Fund -- ESF) and
Greek national funds through the Operational Program ``Education and Lifelong Learning'' of the
National Strategic Reference Framework (NSRF) - Research Funding Program:
``{\sl Thalis. Investing in knowledge society through the European Social Fund}''.}}
\author {Stavros G. Kolliopoulos\thanks{Department of Informatics and
Telecommunications, National and Kapodistrian 
University of Athens, Panepistimiopolis Ilissia, Athens
157 84, Greece; (\texttt{www.di.uoa.gr/}\~{\tt sgk}). Part of this
work was done while visiting the 
  IEOR Department, Columbia University, New York, NY 10027.}   
\and Yannis Moysoglou\thanks{ Corresponding author. Department of Informatics and
Telecommunications, National and Kapodistrian 
University of Athens, Panepistimiopolis Ilissia, Athens
157 84, Greece; (\texttt{gmoys@di.uoa.gr.}) } }
\maketitle

\begin{abstract}
We consider  the following   special case of minimizing makespan.  
%% with assignment constraints. 
A set of jobs $J$ 
and a set of machines $M$ are given. Each job $j \in J$ can be
scheduled on a  machine
from a subset $M_j$ of $M$. The processing time of $j$ is the same on
all  machines in $M_j.$ The jobs  are of two sizes, namely $b$ (big) and
$s$ (small). %% The goal is to schedule the jobs on the machines so as to
%% minimize  the makespan.  
We present  a polynomial-time  algorithm that
approximates the value of the optimal makespan  within a 
factor of $1.883$ and some further improvements  when every job
can be scheduled on at most two machines. 
\end{abstract} 


\section{Introduction}

%Problems of scheduling jobs to machines have been widely studied as they have great practical and theoretical importance.

 The problem we  consider is a special case  of makespan minimization,
 i.e., the problem  of scheduling a set of jobs $J$ on a set of
 machines  
$M$ with the objective
 of  minimizing the  maximum machine load.  In the  most
 general case  of \emph{unrelated} machines,  each job $j \in J$ has a
 processing  time  $p_{ij}$  on  machine  $i \in M.$  An  LP-rounding
 algorithm with an  approximation factor of $2$ along  with a proof of
 $3/2$-hardness, unless ${\sf P} = {\sf
  NP}$, are two
 classic results of \cite{lenstra}. For more than 20 years, 
no progress has
 been made either  on the approximation of the general  case or on the
 lower bound.

The $3/2$  lower bound  holds also for  the case with  {\em assignment
  constraints.} In this setting, each  job $j$ can be scheduled on any
machine from a subset $M_j$ of  $M$. The processing time $p_j$ of $j$ is the
same for any machine in $M_j.$ 
For this case it was recently shown in \cite{svensson} that a strong
LP-relaxation called the configuration LP has an integrality gap of at
most  $33/17$.   The  proof   of  this  exciting   result  is
non-constructive, in  the sense  that it does  not actually  provide a
polynomial-time algorithm  that finds such a solution.  The best known
polynomial-time  algorithm that  computes a  near-optimal  schedule is
still the $2$-approximation algorithm of \cite{lenstra}. A slight improvement to
$2-1/|M|$ was given by \cite{ShchepinV05}. 

The  makespan  minimization  problem  is  one of  the  most  important
problems  in  scheduling.  Advancement  on the  approximation  of  the
unrelated  case  appears as  one  of the  top  $10$  open problems  in
approximation  algorithms in  the  listing of  \cite{sw}.  Due to  the
importance of the problem several special cases have been considered,
where further  restrictions are imposed  either on the sets  $M_j,$ or
on the domain of the processing times. 

 In the case of \emph{graph balancing} each job can be scheduled on at
 most  2  machines. One  can  interpret  this  as  the  following
 problem: we are given an undirected graph with weighted edges and we are asked to
 direct  the edges towards  the nodes  so as  to minimize  the maximum
 weighted in-degree over all nodes. An LP-rounding algorithm achieving
 an approximation factor of  $1.75$ was given in \cite{ebenlendr}. The
 LP  used is  that  of \cite{lenstra}  with  the addition  of a valid set of
  inequalities. The considered  LP has an integrality gap  of $2$ when we
 allow jobs that can be scheduled on $3$ machines \cite{svensson}. The
 case of graph balancing with no parallel edges and with integer edge
 weights had been previously considered in
 \cite{asa}  under  the  name  \emph{graph orientation}.  Among  other
 results         \cite{asa}        gave         a        combinatorial
 $(2-\frac{2}{k+1})$-approximation algorithm when weights are from the 
 set $\{1, k\}.$

The natural case when jobs can be either ``big'' or ``small'', i.e., the processing
times  can  only take one of two values gives
rise to particularly difficult instances.
The unrelated version of graph balancing with only $2$ distinct
edge weights seems to capture a major portion
of the difficulty of the general scheduling problem on unrelated machines.  It was
 recently shown in  \cite{jose} 
that even when  there are $2$ distinct edge weights and 
an  edge   may  have  different  weights  for   each  endpoint,  the
configuration  LP has  an  integrality gap  of  $2$. 
For scheduling with assignment constraints, 
$3/2$-hardness    holds also for 
the case when there are
only $2$  distinct processing time values \cite{lenstra} and  the  integrality  gap  of  the
natural LP of
\cite{lenstra}  is  still $2$ \cite{ebenlendr}.
%% This case is in fact
%% the unrelated version of a restricted version of the problem why
%% consider, namely the unrelated 2-valued graph balancing.

Motivated by the above, 
we study the {\em 2-valued} case of minimizing makespan 
with assignment
constraints, where     jobs are of two sizes,  namely $b$ (big) and
$s$ (small). 
\iffalse ========================================
As mentioned,  the
approximation hardness of the problem is still $3/2,$ unless ${\sf P} = {\sf
  NP}$
(\cite{lenstra},\cite{ebenlendr})  and  the  integrality  gap  of  the
natural LP of
\cite{lenstra}  is  still $2$ \cite{ebenlendr}.  
=================================== \fi
We 
present a  simple  algorithm for the case when the job sizes are
$1$ and $k \in \mathbb{N}$ which approximates the value of the
optimal solution  within a ratio  of $2-\frac{1}{k}$ and then  we show
how to use this algorithm as  a black-box subroutine to get an approximation when
the job  sizes are nonnegative  real numbers. Combining  our algorithm
with the non-constructive result of \cite{svensson} we get an improved
non-constructive approximation guarantee  of $1.883$ for the $2$-valued
case of  makespan with assignment constraints. The  same algorithm can
be used to obtain an efficient $1.652$ approximation for the $2$-valued
graph balancing, where additionally for every $j$ we have $|M_j| \leq 2$, 
improving and generalizing the result of
\cite{asa}. Note  that even
for the 
2-valued graph
balancing 
it remains {\sf NP}-hard to compute a better than $3/2$-approximate
solution 
\cite{asa, ebenlendr}.

%
%\begin{table}
%\begin{center}
%\begin{tabular}{ l | c | r | }
%\hline
%Problem&Known Approximation&Our Contribution\\
%\hline
% \{1,k\} graph balancing & 1.75 \cite{ebenlendr},2-1/k \cite{asa}& 1.5 \\
 % 2-value graph balancing & 1.75\cite{ebenlendr} & 1.66 \\
 % 2-value restricted makespan & 1.94,1.66+s \cite{svensson}& 1.88 \\
%\hline
%\end{tabular}
%\end{center}
%\caption{Known approximation factors for the problems we consinder.}
%\end{table}

\section{An algorithm for the case with job sizes from the set $\{ 1, k \}$ }

This section describes a subroutine which will be used in our
algorithms. 
Apart from a few details, the network construction and the rounding
argument follow from \cite{Kleinberg96a}. 

Similar to \cite{lenstra}, our algorithm takes as a parameter an estimation $T$ of the makespan of the optimal solution. Using binary search, we find the smallest $T$ for which our algorithm returns a (possibly fractional) solution. As we will show, that $T$ is a lower bound to the makespan of the optimal solution.

Here and in the following sections we make the assumption that the optimal makespan is less than twice the size of the biggest job $b$. This is w.l.o.g. since the algorithm 
of \cite{lenstra} returns a solution with makespan at most $T^*+b,$
where  $T^*$ is the optimum of the natural LP relaxation for the
problem.  This is formalized in the following lemma. 

\begin{lemma} \cite{lenstra} \label{lemma:add}
If the optimal makespan to an instance of the problem is $T_ {Opt} \geq 2b$, where $b$ is the biggest job size, then there is $3/2$-approximation algorithm. 
\end{lemma}
\iffalse =========================
\begin{proof}
The rounding algorithm of \cite{lenstra} returns a solution with makespan at most $T_ {Opt}+b$. This is clearly a $3/2$ approximate solution.
\end{proof}
====================== \fi 

\begin{figure}

\begin{center}
\scalebox{.4}{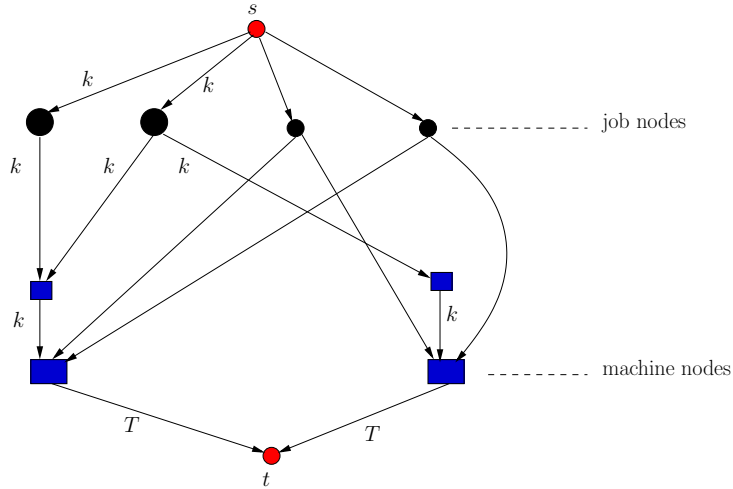}
\end{center}

\caption{Example network construction for an instance with $4$ jobs and $2$ 
machines. Edges have unit capacity unless indicated otherwise.} \label{fig:graph}
\end{figure}

The algorithm we describe is used for instances where the sizes of the jobs are integers, either 1 or $k>1$.
Given  such an  instance  of   the  problem,  we  construct  the  following
multi-level  network. On  the  first  level we  have  a single  source
$s$. On the  next level we have one node $n_j$ for each  job $j$. For
each such  node $n_j$  we have a  directed edge  from $s$ to  $n_j$ of
capacity equal  to the size  of job $j$.  On the level after  that, we
have a node $v_{i,b}$ for each  machine $i.$ This type of nodes is not
actually representing the  set of machines. Their purpose  is to limit
the amount of flow from big jobs that may enter the machine nodes of
the next level, enforcing  the valid inequality of \cite{ebenlendr} in
our solution.  
%% We note  that apart  from this detail,  the construction we employ  is
%% fairly standard in the literature. 
We have a a directed  edge of capacity $k$  from a node
$n_{j}$ to a node  $v_{i,b}$ iff $j$ is a big job that can be scheduled
to machine  $i$. On  the next level  we have  one node $m_i$  for each
machine $i$.  We have a directed  edge of capacity $1$  from each node
$n_{j}$, where job $j$  is a small job, to a node  $m_i$ iff job $j$ can
be scheduled  on $i$. We also  have one directed edge  of capacity $k$
from $v_{i,b}$  to $m_i$.  We connect each  machine node $m_i$  to the
sink $t$ at  the last level using edges of capacity  $T$, where $T$ is
our  estimation for  the makespan.  See Fig.~\ref{fig:graph} 
for an example. Observe that  while the  flow from
small job nodes is routed directly to the machine nodes, the flow from
the big  job nodes is routed  through the nodes $v_{i,b}$  and then to
the machine nodes.  Thus at most $k$ units of flow  from big job nodes
enter each machine node.

A feasible flow solution is one that sends a 
total amount of flow  equal to the  sum of the
processing times of the jobs. Using binary search we find the minimum
value of $T$ for which such a feasible solution exists. 
This is a lower bound on the optimum. To see why this is true,
consider a solution $S$ to the problem instance with makespan $T$. We
construct the following network flow solution for the network
corresponding to the instance with estimation $T$ or higher: for each job $j$,
we send from $s$ to $n_j$ flow equal to the size of $j$. If $j$ is a
small job assigned to machine $i$ by $S$, then we send $1$ unit of
flow from $n_j$ to $m_i$. If $j$ is a big job assigned to machine $i$
by $S$, then we send $k$ units of flow from $n_j$ to $v_{i,b}$ and
then $k$ units to $m_i$ from $v_{i,b}$ (there is at most one such big
job for each machine -- recall that because of  Lemma~\ref{lemma:add} we consider only instances with $T< 2b$). We send from each node $m_i$ to the sink $t$ all amount of flow that $m_i$ has received, which is at most $T$.
 We
interpret the  network solution naturally as  a fractional assignment:
if a fraction $f$  of the flow that source $s$ sends  to node $n_j$ is
routed through the machine node $m_i$ then we assign a fraction $f$ of
job $j$ to machine $i$.
 
\iffalse    ==== %% redundant
We  are only  interested in  instances of  the problem  for  which the
network  has a  feasible solution  for  some value  of the  parameter
$T$. If an instance has no feasible solution for any value of $T$, it
 it is necessary to have at least 2 big jobs scheduled
on  the same  machine. Obtaining  a  $3/2$ approximation  for such  an
instance is trivial via the rounding technique of \cite{lenstra}.
\fi

\iffalse =========================== condensed ==============
After  solving the  network flow  problem and  obtaining  a fractional
assignment, we  can easily produce the  following integral assignment:
the small jobs are already  integrally assigned to machines because of
flow integrality.  Observe that each  machine has a total  fraction of
big jobs of at most $1$ assigned to it, due to the capacity $k$ of the
edge that connects node $v_{i,b}$ to node $m_i$. Moreover the fraction
of a  big job that is  assigned to a  machine is a multiple  of $1/k$,
again due to  flow integrality. This guarantees that if  a big job $j$
has a non zero fraction of it  assigned to a machine $i$, then $i$ has
a fraction of $j$ of at least $1/k$.

The above gives us the following :

\begin{lemma}
Let $S$ be  the fractional assignment that corresponds  to the network
flow solution  of an instance of  the problem. Then for  every big job
$j$ that has  a non zero fractional assignment  $x_{j,i}$ to a machine
$i$, $x_{j,i}\geq 1/k$.
\end{lemma}
================== ================ \fi 

After  solving  the  network  flow  problem  we  obtain  a  fractional
assignment $S$ in which small jobs are integrally assigned and for  every big job
$j$ that has a non-zero fraction $x_{ij}$ assigned to machine
$i \in M_j,$ $x_{ij}\geq 1/k$. In particular by flow integrality, if
$x_{ij} >0,$ then $x_{ij}$ will be an integral multiple of $1/k.$

The next lemma shows that we can assign each big job to a machine that
has a fraction of  at least $1/k$ of it, so that  each machine gets at
most one big job. 
%% Note that, if a machine has a non-zero fraction of a big job, that
%% fraction is an integral multiple of $1/k$ from flow integrality. 
The  proof of the lemma is easy and is omitted. 

\begin{lemma} \label{lemma:easy} 
Let $J_b$ be the  set of big jobs and $S(J_b)$ be  the set of machines
that have a non-zero fraction of  a big job. We can find in polynomial
time a feasible schedule of the jobs of $J_b$ to the machines of $S(J_b)$ 
so that at most $1$ big job is assigned to each machine.
\end{lemma}

\iffalse =========== omit proof ==================
\begin{proof}
We form  the following  bipartite graph $G(V_1,V_2,E)$  : on  one side
$V_1$ of $G$ we have one node for each big job $j \in J_b$, and on the
other side $V_2$ we have one node for each machine $m \in S(J_b)$. The
set of edges $E$ is defined as  follows. We have an edge from each big
job node to each machine node  iff the corresponding machine has a non
zero fraction  of that job. Now let  us consider a set  of nodes $V'_1
\subseteq V_1$  and let  $\Gamma (V'_1) \subseteq  V_2$ be the  set of
neighbor nodes of  $V'_1$, that is $\Gamma (V'_1)$  is exactly the set
of  nodes that are  connected to  some node  of $V'_1$.  Then $|\Gamma
(V'_1)|  \geq |V'_1|$. Let  assume otherwise  that $|\Gamma  (V'_1)| <
|V'_1|$. Since  our network solution  sends $k$ units of  flow through
any node that corresponds to a big job $j$, the total fractions of $j$
that are assigned to machines  in our fractional assignment $S$ sum up
to $1$.  Thus the  total fraction of  the big  jobs in $V'_1$  that is
assigned  to  the  machines  of  $\Gamma (V'_1)$  is  $|V'_1|$.  Since
$|\Gamma  (V'_1)| <  |V'_1|$ there  is at  least one  machine  that is
assigned a total fraction of big jobs greater than $1$, which violates
the property of our fractional assignment discussed above.

Thus the conditions  of Hall's theorem for bipartite  matching hold, so
we  can in  polynomial time  find such  an assignment  of big  jobs to
machines, with each machine getting at most one big job.
\end{proof}
================= End omit proof ============ \fi 

The value of the optimal solution is at least $k,$
the size of a big job. By Lemma~\ref{lemma:easy} 
in our solution we increased the load of a
machine by at most $k-1.$ This results in a $2-1/k$ approximation.

\begin{proposition} \label{propos}
The solution we get for an instance of the $2$-valued makespan problem
with job  sizes from the  set $\{ 1,k \}$ by rounding  the fractional
assignment  $S$  resulting from  the  network  flow  solution, has  an
approximation ratio of $2-1/k$.
\end{proposition}

\section{The Case With Arbitrary Job Sizes } \label{sec:int}

In this section we will use the algorithm of Proposition~\ref{propos}  to
design an approximation algorithm for  the more general case where the
two job sizes are arbitrary  nonnegative real numbers. Without loss of
generality we assume that $b=1$. We can easily normalize the job sizes
by dividing them by $b$. We also assume that the size of small jobs is
$1/\alpha $ for  some $\alpha > 1$. We can reduce  this problem to the
previous one  by changing the size  of small jobs  to either $1/\lceil
\alpha  \rceil$ or  $1/\lfloor \alpha  \rfloor$ and  use  the previous
algorithm, since  in the modified instance 
the size of the big  jobs is an integer  multiple of the
size of the small jobs. We will use the resulting assignment as a solution
to  our  original  instance. Below  we  prove  by  doing so  that  the
approximation ratio  is bounded by certain expressions.  Later we will
combine those expressions with the result of \cite{svensson} to get an
improved non-constructive approximation guarantee.

\begin{definition}
Let $I$ be the 
 original instance of our problem with job sizes in $\{1/\alpha ,1
 \}$, $\alpha > 1.$ Let $I_1$ be the instance where we change the small job sizes to $1/\lceil \alpha  \rceil $ and $I_2$ be the instance resulting from changing the small job sizes to $1/\lfloor \alpha  \rfloor$. Additionally let $f_1=\lceil \alpha  \rceil /\alpha  $ be the factor by which the small jobs in $I$ are greater than the small  jobs in $I_1$, and $f_2=\lfloor \alpha  \rfloor /\alpha $ be the factor by which the small jobs in $I$ are smaller than the small jobs in $I_2$.
\end{definition}

We proceed by bounding the approximation ratio in the case we use the solution of instance $I_1$ as a solution to $I$. We focus on the load of some machine $i$. Let $Opt_1$ be the  cost of the optimal solution of instance $I_1$ and $Opt$ be the cost of the optimal solution for the original instance $I$. In the fractional assignment resulting from the solution of the network flow for instance $I_1$, let $B_1$ be the total fraction of big jobs that are assigned to $i$. If $B_1 > 0$ then $B_1 \geq 1/\lceil \alpha  \rceil$ by flow integrality. Then the load of $i$ due to small jobs is at most $Opt_1 - B_1$. After the rounding the load of $i$ is at most $1+Opt_1-B_1$ where the load due to big jobs is $1$ and the load due to small jobs is $Opt_1-B_1$. If $B_1=0$, only small jobs are assigned to machine $i$, then the load of $i$ does not increase during the rounding and thus the total load of $i$ is at most $Opt_1$.

If we use the forementioned solution of $I_1$ as a solution to $I$, since the small
jobs in $I$ are $f_1$ times greater, the load of a machine $i$ that is assigned a big job 
will be  at most
$1+(Opt_1-B_1) f_1$.  Also we have that $Opt \geq Opt_1$. So, regarding the constructed solution for the original instance $I$, the ratio of the load of $i$ to the optimum makespan is $\frac{1+(Opt_1-B_1) f_1}{Opt} \leq \frac{1+(Opt_1-B_1) f_1}{Opt_1}$. Since $B_1 \geq 1/\lceil \alpha  \rceil $ and $Opt_1 \geq 1$ we upper bound the former ratio once more (setting $B_1 =1/\lceil \alpha  \rceil $ and $Opt_1 = 1$) by $1+f_1-1/a$ . If $i$ is assigned only small jobs, then the corresponding load is at most $Opt_1f_1$ in instance $I$ and the corresponding ratio is at most $f_1$. Clearly $1+f_1-1/\alpha \ge f_1$, since $\alpha >1$. Thus we can bound the approximation ratio achieved using the solution of $I_1$ by the ratio of the first case: $1+f_1-1/\alpha$.

Now we will make a similar analysis for the case we use the solution
of $I_2$ as a solution to $I$. Let $Opt_2$ be the value of the optimal
solution of $I_2$ ($Opt_2 \geq 1$). We once again focus on the load of a
single machine $i$. Using the same reasoning as above, if $i$ was assigned a non-zero fraction $B_2$ of  big jobs, we have $B_2 \geq 1/\lfloor \alpha  \rfloor$ by flow integrality. The load of $i$
after rounding the fractional assignment is at most $1+(Opt_2-B_2)$. The
small jobs of $I$ have size $f_2$ times the size they have in
$I_2$. So the load of $i$ for instance $I$ is at most
$1+(Opt_2-B_2)f_2$. As for the value  $Opt$ of the optimal solution of $I$, we know that $Opt \geq Opt_2 f_2$ (since the optimal solution to $I$ with cost $Opt$ induces a solution to $I_2$ which is at most $Opt/f_2$ ). Thus the ratio of the contructed solution to the optimal cost is $\frac{1+(Opt_2-B_2)  f_2}{Opt} \leq \frac{1+(Opt_2-B_2) f_2}{Opt_2 f_2}$. Like before, we upper bound the former expression by setting $Opt_2=1$ and $B_2=1/\lfloor \alpha  \rfloor$ and we get $1/f_2+1-1/\lfloor \alpha  \rfloor$. If $i$ is assigned only small jobs, then the load does not increase during the rounding and is at most $Opt_2$ regarding instance $I_2$. The corresponding load of $i$ in the solution for instance $I$ is at most $Opt_2f_2$. The ratio of this case is $Opt_2f_2/Opt \leq Opt_2f_2/Opt_2f_2=1$. Since $\alpha >1$, $1/f_2+1-1/\lfloor \alpha  \rfloor= \alpha/\lfloor \alpha  \rfloor+1-1/\lfloor \alpha  \rfloor > 1$. So, once again, the expression that bounds the approximation ratio achieved is $1/f_2+1-1/\lfloor \alpha  \rfloor$.

We use the solution of the instance that achieves the minimum approximation ratio. We have thus proved the following theorem:

\begin{theorem}
For any instance of the 2-valued makespan with assignment constraints
and with job sizes in $\{ 1, 1/\alpha  \}$, $\alpha  >1,$ we can find in polynomial time a solution which has an approximation ratio $\min \{1+f_1-1/\alpha,1/f_2+1-1/\lfloor \alpha  \rfloor \}$
\end{theorem}

The approximation we have achieved so far depends on $\alpha$. We can calculate  
 the worst case approximation given by the above theorem for interval $(n,n+1)$ which contains $\alpha$, by setting the two expressions to be equal, since one is decreasing  and the other is increasing in $\alpha$, for a given interval $(n,n+1)$.

According to the above, the worst approximation achieved for $\alpha \in (n,n+1)$ is for the value of $\alpha$ for which:

\begin{center}
$1+f_1-1/\alpha=1/f_2+1-1/\lfloor \alpha  \rfloor \Leftrightarrow $\\
$1+\frac{n+1}{\alpha}-1/\alpha=\frac{\alpha}{n}+1-1/n \Leftrightarrow $\\
$\alpha^2-\alpha-n^2=0$\\
\end{center}

The solution is  $\alpha \in 
\{ \frac{1-\sqrt{1+4n^2}}{2}, \frac{1+\sqrt{1+4n^2}}{2}\} \cap (n,n+1)$.

 The above calculation  gives an approximation guarantee of $1.883$ for values of $\alpha $ up to $5$. For values of $\alpha > 5$ the small job size is $1/\alpha  < 0.2$ and therefore it is preferable to use the following:

\begin{theorem}
\cite{svensson} If an instance of the scheduling problem only has job sizes $s \geq 0$ and $1$, then the configuration LP has integrality gap at most $5/3+s$.
\end{theorem}

The above theorem was actually proved in \cite{svensson} with the assumption that $T=1$. The proof, however, can be easily generalized to the case where $T<2$ by making some minor changes. The main idea is changing $1$ to $T$ wherever the proof refers to the makespan (i.e., the integrality gap becomes $(T+2/3+s)/T$).
Note that for $\alpha  > 5$ the approximation of the above theorem is at most $5/3+0.2<1.883$. So the following has been proved:

\begin{theorem}
We can estimate the optimal makespan of a 2-valued  instance with  assignment constraints and with jobs of two sizes within a factor of $1.883$ in polynomial time.
\end{theorem}

\section{2-valued case of graph balancing}

In this section we present a $1.652$-approximation for graph balancing instances, where the edge weights belong to the set $\{ b,s \}$. Our proof relies on a modification of the proof for the more general case of 2-valued makespan with assignment constraints. We will first prove a tight $3/2$ approximation when $b=k, s=1, k \in \mathbb{N}$.

Let us  consider the multi-level  network defined above for  this case
where the graph edges take  the role of jobs and the nodes  take the role of
machines. The key difference here is  that for each job node there are
at  most $2$  flow  paths  to the  set  of machine  nodes.  So in  the
fractional solution resulting from the flow solution, for each big job
$j$ there are at most $2$  machines that have nonzero fraction of that
job. Moreover  for each  big job  either there is  one machine  with a
fraction greater  than $1/2$  of $j$, or  there are $2$  machines with
exactly a $1/2$  fraction of $j$ each. This is  due to the constraint
enforced by nodes $v_{i,b}$ that each machine may take a total fraction of
big jobs of at most $1$. The small jobs are assigned integrally due to
integrality of flow.

Now we show  how to round the fractional assignment:  for each big job
$j$ for  which there  is a  machine $i$ with  a fraction  greater than
$1/2$ of $j$  assigned to it, assign  $j$ to $i$. As for  the big jobs
that are assigned to $2$ machines with a fraction of exactly $1/2$, we
consider the graph induced by the corresponding edges (only for those big jobs) and the nodes(machines) covered by those edges. Since
each  node has  degree  at most  $2$ (otherwise a machine is assigned at least 3/2 big jobs, which cannot happen),  the graph  is  a collection  of
disjoint paths  and cycles. It is  easy to find a  $1-1$ assignment of
the edges  to the  nodes (i.e., make  a clockwise assignment  on cycles,
direct each path arbitrary and assign each edge to its head).

Note that in the resulting assignment, each node-machine which had a fraction at least $1/2$ of some big edge-job may end up taking the whole edge-job. We have increased the cost of the fractional solution by $1/2$ at most and since the cost of the optimal solution is at least $1$, we have achieved a $3/2$ approximation, matching the lower bound.

\begin{theorem}
For instances of graph balancing in which %the edge weights belong to the set%
$b=k, s=1, k \in \mathbb{N},$ we can  find in polynomial time a $3/2$-approximate solution.
\end{theorem}

As in the previous section, we  use the above algorithm as a black box
to solve the case with edge weights in $\{ b,s\}$ or w.l.o.g. in $\{ 1,1/\alpha \}.$
Following the exact same  argumentation, we can reduce this case
to the  previous one,  by rounding  the small job  size to  either $1/
\lceil  \alpha  \rceil$  or   to  $1/  \lfloor  \alpha  \rfloor$. For both cases, the worst approximation ratio arise again from the case where machine $i$ has a non-zero fraction of big jobs and received a whole big job after the rounding. Now we know that $i$ had at least $1/2$ fraction of some big job. The
expressions  giving   the  approximation   in  each  case   are 
$1+f_1/2$  and $1/f_2+1/2$ respectively and are obtained in the same
manner as in Section~\ref{sec:int}. 
%% Note that when $\alpha \in (1,2)$, the expression are exactly the same as those of the %%previous section ($B_1,B_2 \geq 1/2$), achieving a better than $1,653$ approximation. So %%assume $\alpha \ge 2$.
 Once again we  balance the  expressions in  each
interval $(n,n+1)$ getting the following:

\begin{center}
$1+f_1/2=1/f_2+1/2 \Leftrightarrow$\\
$2\alpha^2-n\alpha-(n+1)n=0$
\end{center}

Assume $\alpha \ge 2$. Solving the above equation for the intervals, the worst approximation achieved is when $\alpha \in (2,3)$, which is less than $1.652$.

When $\alpha \in (1,2)$  we use a different approach: in this case
$s=1/\alpha \ge 1/2$. If $Opt=1$ then it is obvious that we can find
an optimal solution via bipartite maching. If $Opt>1$, then we know
that $Opt \geq 2s$, since there must be a machine with at least 2 jobs
assigned in each solution. Consider the case $s\geq 0.65$. If $Opt > 2s$ then $Opt \geq 1+s \geq 1.65$ (since we have either $1$ big job and $1$ small in some machine or we have more than 3 small jobs in some machine) and the
algorithm of \cite{lenstra} gives a $\frac{1+1.65}{1.65} < 1.652$
approximation. If $Opt=2s$ using, e.g., the LP and the cycle canceling
of \cite{ebenlendr} we get a fractional solution of cost  at most $Opt,$  for which the corresponding graph induced on the fractionally assigned nodes (machines) is a forest. Consider a tree $t$, and further consider a leaf node $l$. The load of the integrally assigned jobs of $l$  can either be $1$ or $s$ ($l$ cannot have 2 integrally assigned jobs and one fractionally since $Opt=2s$). We assign the fractional job to $l$, resulting to a total load of at most $2$. We do the same in a bottom-up manner for each node of the tree. The resulting approximation is $2/2s$ which is less than $1.652$ for $s \geq 0.65$. If $s<0.65$ then the reduction to $I_1$ done previously gives an approximation of at most $1+f_1/2 =1+\frac{2}{2a}=1+s < 1.652$. Note that, for all the above, we do not need  to know the value of $Opt$, we just keep the best solution of the mentioned approaches.

\begin{theorem}
For instances of graph balancing in which the edge weights belong to
the set $\{ b,s \}$ we 
can compute in polynomial time  a $1.652$-approximate solution.
\end{theorem}

\bibliographystyle{plain}

%% \bibliography{biblio,/Users/stavros/papers/bibliography}

\end{document}